\documentclass[11pt,a4paper,pdftex]{article}
\pdfoutput=1
\usepackage{jcappub}

\newcommand{\be}{\begin{eqnarray}}
\newcommand{\ee}{\end{eqnarray}}

\title{Testing SgrA$^*$ with the spectrum of its accretion structure}

\author{Nan Lin$^a$, Zilong Li$^a$, Jake Arthur$^{b,a}$, Rachel Asquith$^{b,a}$\\
and Cosimo Bambi$^{a,1}$ 
\note{Corresponding author}}

\affiliation{$^a$ Center for Field Theory and Particle Physics and Department of Physics,\\
Fudan University, 220 Handan Road, 200433 Shanghai, China}

\affiliation{$^b$ School of Physics \& Astronomy, The University of Nottingham,\\
University Park, Nottingham NG7 2RD, United Kingdom}

\emailAdd{linnan8957@gmail.com}
\emailAdd{zilongli@fudan.edu.cn}
\emailAdd{ppyjta@nottingham.ac.uk}
\emailAdd{ppyra@nottingham.ac.uk}
\emailAdd{bambi@fudan.edu.cn}

\abstract{SgrA$^*$ is the supermassive black hole candidate at the center of the Galaxy and an ideal laboratory to test general relativity. Following previous work by other authors, we use the Polish doughnut model to describe an optically thin and constant angular momentum ion torus in hydrodynamical equilibrium and model the accretion structure around SgrA$^*$. The radiation mechanisms are bremsstrahlung, synchrotron emission, and inverse Compton scattering. We compute the spectrum as seen by a distant observer in Kerr and non-Kerr spacetimes and we study how an accurate measurement can constrain possible deviations form the Kerr solution. As in the case of emission from a thin accretion disk, we find a substantial degeneracy between the determination of the spin and of possible deviations from the Kerr geometry, even when the parameters of the ion torus are fixed. This means that this technique cannot independently test the nature of SgrA$^*$ even in the presence of good data and with the systematics under control. However, it might do it in combination with other measurements (black hole shadow, radio pulsar, etc.). }


\begin{document}

\maketitle


\section{Introduction}

SgrA$^*$ is the radio source associated to the supermassive black hole (BH) candidate at the center of our Galaxy~\cite{sgra}. Its mass is around 4~million Solar masses and its distance from us is about 8~kpc. This makes SgrA$^*$ a peculiar source and an ideal laboratory to test general relativity. It is much closer than any other supermassive BH candidate and much heavier than Galactic stellar-mass BH candidates in X-ray binaries. It can thus be studied with observations not feasible in other cases. For instance, its apparent angular size on the sky is expected to be about 50~$\mu$as and we may be able to image its shadow in the near future~\cite{sh1,sh2}.

Astrophysical BH candidates are so called because they can be naturally interpreted as the Kerr BHs predicted by general relativity, but direct observational evidence is still lacking~\cite{narayan,cb1,cb2}. Today we have robust measurements of the masses of these objects. This is enough to conclude that stellar-mass BH candidates in X-ray binaries are too heavy to be neutron stars~\cite{rr} and that supermassive BH candidates in galactic nuclei are too heavy, compact, and old to be clusters of neutron stars~\cite{maoz}. After that, we have to rely on the validity of general relativity, even if the theory has only been tested in weak gravitational fields, mainly with Solar System experiments and radio pulsar data. In general relativity and under a set of reasonable assumptions about the matter content and the spacetime structure, the final product of a gravitational collapse is a BH~\cite{daniele,luca}. The only uncharged BH solution in 4D is the Kerr metric and therefore BH candidates must be Kerr BHs if general relativity is correct and they may not be Kerr BHs only in presence of new physics.

In the past few years, there have been significant efforts to figure out how to test the nature of BH candidates in order to confirm the Kerr paradigm. The study of the thermal spectrum of geometrically thin disks and the analysis of the iron K$\alpha$ line are the two main techniques capable of probing the spacetime geometry around these objects and test the Kerr metric~\cite{cfm1,cfm2,iron1,iron2}. However, the thermal spectrum of a thin disk has a very simple shape and there is a degeneracy between the estimate of the spin and of possible deviations from the Kerr background. Eventually, even large deviations from the Kerr solutions are allowed~\cite{cfm3,cfm4}. The profile of the iron line has a more complicated structure and it is potentially a more powerful tool to test the Kerr metric~\cite{jjc1,jjc2}. However, high quality data would be necessary. With the available observations, we can conclude that BH candidates cannot be some types of compact dark stars or wormholes~\cite{e1,e2}, because the iron lines of these objects would be qualitatively different from those observed, but we still cannot exclude large deviations from the Kerr geometry~\cite{jjc2}.

Since a number of studies has shown that the main obstacle to test the Kerr paradigm is the strong correlation between the spin and possible non-Kerr features, it sounds natural to try to break the degeneracy with a combination of different measurements of the same source~\cite{deg1,deg2,deg3,deg4,polar}. With this spirit, SgrA$^*$ may soon become one of the best objects to test the Kerr metric. While there are currently no measurements to probe the spacetime close to this object, there are strong motivations to expect that near future facilities will be able to provide a number of different data. The detection of the boundary of the BH shadow can provide a measurement of the apparent photon capture radius~\cite{vlbi1,vlbi2}, which only depends on the metric of the spacetime~\cite{o1,o2,o3,o4,o5,ae1,ae2,ae3,lanzhou,aaa1,aaa2}. The observation of blobs of plasma orbiting near the innermost stable circular orbit can measure the orbital frequency and effects related to strong gravitational lensing~\cite{blob0,blob1,blob2,blob3}. The observation of a radio pulsar in a close orbit, but also of ordinary stars, can provide a measurement of the spin of SgrA$^*$~\cite{pulsar,ss1,ss2}. The combination of different measurements seem to be the right strategy to break the degeneracy between an estimate of the spin and possible non-Kerr features, and SgrA$^*$ may be the best source to get several independent observations to combine together~\cite{comb1,comb2}.

In this paper, we want to explore one more approach to test the Kerr metric with SgrA$^*$. The accretion structure around SgrA$^*$ seems to be a radiatively inefficient advection dominated accretion flow~\cite{adaf}. As proposed in Refs.~\cite{straub,vincent}, such an accretion structure may be described by an optically thin and constant angular momentum ion torus within the Polish doughnut model~\cite{pd1,pd2} and a 2-temperature plasma in which the emission mechanisms are bremsstrahlung, synchrotron radiation, and inverse Compton scattering. Since the structure of the torus is determined by the spacetime geometry, the comparison between theoretical predictions  with observations can potentially constrain the metric around SgrA$^*$. In this explorative work, we compute the spectrum of ion tori in Kerr and non-Kerr backgrounds and pay some attention to the correlation among the spin parameter of the central object, possible deviations from the Kerr solution, and the fluid angular momentum. These three parameters are related to the spacetime metric. However, the spectrum also depends on 5 other parameters. Here we do not study the correlation among all the parameters in the model.

Even when the parameters of the ion torus are fixed, the spectrum is degenerate with respect to the spin and the deformation parameters and it is impossible to test the Kerr metric without other data. The situation is quite similar to that of the measurement of the thermal spectrum of a thin disk of a stellar-mass BH candidate in a binary. When the fluid angular momentum is also a free parameter to be determined by the fit, the constraining power of the spectrum is substantially the same. In order to test SgrA$^*$, it is necessary to combine different observations. The constraints from the spectrum of the accretion structure might be used, for instance, in combination with those from the observation of the BH shadow, from the detection of a pulsar/stars in close orbits, and from the observations of blobs of plasma orbiting near the innermost stable circular orbit.

The content of the paper is as follows. In Section~\ref{s-2}, we review the Polish doughnut model in non-Kerr backgrounds. In Section~\ref{s-3}, we summarize the ion torus proposal of Ref.~\cite{straub}. In Section~\ref{s-4}, we show the results of our simulations in Kerr and non-Kerr spacetimes and we study the correlation between the estimate of the spin and of possible deviations from the Kerr geometry. Summary and conclusions are reported in Section~\ref{s-5}. Throughout the paper, we employ units in which $G_{\rm N} = c = 1$ and the convention of a metric with signature $(-+++)$.

\section{Polish doughnut model in non-Kerr spacetimes \label{s-2}}

Since we want to test the Kerr metric, we consider a background more general than the Kerr solution and that includes the Kerr solution as special case. In addition to the mass $M$ and the spin parameter $a_* = a/M = J/M^2$, where $J$ is the BH spin angular momentum, the metric has one or more ``deformation parameters'', which are used to quantify possible deviations from the Kerr geometry. The idea is to compute the spectrum of the ion torus in this more general background and see whether observations can potentially measure both the spin and the deformation parameter. The latter vanishes in the Kerr spacetime, but here it is assumed a free parameter to be inferred by observations. If the data require a vanishing deformation parameter, then the Kerr paradigm is confirmed. As a non-Kerr background, we consider the Johannsen-Psaltis metric~\cite{jp-m}. The simplest non-Kerr model has only one non-vanishing deformation parameter $\epsilon_3$. In Boyer-Lindquist coordinates, the line element reads
\be\label{gmn}
ds^2 &=& - \left(1 - \frac{2 M r}{\Sigma}\right) \left(1 + h\right) dt^2
- \frac{4 a M r \sin^2\theta}{\Sigma} \left(1 + h\right) dtd\phi
+ \frac{\Sigma \left(1 + h\right)}{\Delta + a^2 h \sin^2 \theta} dr^2
+ \nonumber\\ &&
+ \Sigma d\theta^2 + \left[ \left(r^2 + a^2 +
\frac{2 a^2 M r \sin^2\theta}{\Sigma}\right) \sin^2\theta +
\frac{a^2 (\Sigma + 2 M r) \sin^4\theta}{\Sigma} h \right] d\phi^2 \, ,
\ee
where $\Sigma = r^2 + a^2 \cos^2\theta$, $\Delta = r^2 - 2 M r + a^2$, and
\be
h = \frac{\epsilon_3 M^3 r}{\Sigma^2} \, .
\ee
The compact object is more prolate (oblate) than a Kerr BH with the same spin
for $\epsilon_3 > 0$ ($\epsilon_3 < 0$); when $\epsilon_3 = 0$, we recover the 
Kerr solution.

The Polish doughnut model describes non-self-gravitating disks when the gas pressure cannot be neglected~\cite{pd1,pd2}. This makes the disk geometrically thick and the particles of the gas do not follow the geodesics of the spacetime. The model only requires that the spacetime is stationary and axisymmetric, and therefore its extension to non-Kerr backgrounds is straightforward, but new phenomena may show up~\cite{zilong}. The accretion flow is modeled as a perfect fluid with purely azimuthal velocity, so its energy-momentum tensor and 4-velocity are
\be
T^{\mu\nu} = (\rho + P)u^\mu u^\nu + g^{\mu\nu} P \, , \quad
u^\mu = \left( u^t, 0 , 0 , u^\phi \right) \, ,
\ee
where $\rho$ and $P$ are, respectively, the energy density and the pressure of the gas. In what follows, we assume a polytropic equation of state
\be
P = K \rho^{1 + 1/n} \, ,
\ee
where $K$ is the so-called polytropic constant and $n$ is the polytropic index. The specific energy of the fluid element, $-u_t$, its angular velocity, $\Omega = u^\phi/u^t$, and its angular momentum per unit energy, $l = -u_\phi/u_t$, are
\be
u_t = - \sqrt{\frac{g^2_{t\phi} - g_{tt}g_{\phi\phi}}{g_{\phi\phi} +
2 l g_{t\phi} + l^2 g_{tt}}} \, , \quad
\Omega = - \frac{l g_{tt} + g_{t\phi}}{l g_{t\phi}
+ g_{\phi\phi}} \, , \quad
l = - \frac{g_{t\phi} + \Omega g_{\phi\phi}}{g_{tt}
+ \Omega g_{t\phi}} \, .
\ee
For a stationary and axisymmetric flow in a stationary and axisymmetric spacetime, $l$ is conserved~\cite{pd1}. In the rest of this paper, we consider the simplest model with $l=const.$~\cite{pd2}.

It is useful to introduce the potential $W$
\be
W(r,\theta) = \frac{1}{2} \ln \left[- \frac{g_{tt} 
+ 2 \Omega g_{t\phi} + \Omega^2 g_{\phi\phi}}{\left(g_{tt} 
+ \Omega g_{t\phi}\right)^2}\right] \, .
\ee
The ``center'' of the accretion disk corresponds to the maximum of $P$ and of $W$, say $W=W_c$. The surface of the torus is at some equipotential surface $W=W_s$ where $P=0$. Accretion is possible only when the fluid surface has some cusp: like the cusp at the $L_1$ Lagrange point in a close binary, the accreting gas can fill out the Roche lobe and then be transferred to the compact object. A cusp only exists when the fluid angular momentum $l$ is between a maximum and a minimum value, say $l_{\rm max}$ and $l_{\rm min}$. If $l>l_{\rm max}$, the angular momentum is too high and the fluid cannot approach the central object. If $l<l_{\rm min}$, the angular momentum is too low and there is no stationary accretion flow: all the gas is quickly swallowed by the BH. In the Kerr metric and in many other backgrounds, $l_{\rm max}$ corresponds to the angular momentum per unit energy of the marginally bound equatorial circular orbit, while $l_{\rm min}$ is the angular momentum per unit energy of the marginally stable equatorial circular orbit. In some Johannsen-Psaltis spacetimes, there is no marginally bound equatorial circular orbit, but it is still possible to define $l_{\rm max}$, see Ref.~\cite{zilong} for more details. We just note that, in those backgrounds in which there is no marginally bound equatorial circular orbit, the accretion disk has two cusps, one above and one below the equatorial plane of the system. Instead of $l$ and $W$, we can use the dimensionless parameters
\be
\lambda = \frac{l - l_{\rm min}}{l_{\rm max} - l_{\rm min}} \, , \quad
\omega(r, \theta) = \frac{W(r, \theta) - W_s}{W_c - W_s} \, .
\ee
$\lambda$ can range from 0 to 1 by definition. The same is true for $\omega$ in the region occupied by the gas.

\section{Plasma model \label{s-3}}

The accretion structure around SgrA$^*$ seems to be radiatively inefficient and advection dominated~\cite{adaf}. Following Refs.~\cite{straub,vincent}, we employ the Polish doughnut model (reviewed in the previous section) to describe the macroscopic structure of the accretion disk and a gas pressure dominated optically thin 2-temperature plasma model for the microscopic physics. In this section, we just report the results, while more details can be found in Refs.~\cite{straub,vincent}. The model has 8 parameters: 2 related to the spacetime metric, 1 parameter associated to the position of the observer, and 5 parameters for the accretion flow and its microphysics:
\be
\begin{array}{lll}
\text{background metric} \hspace{0.5cm} & \text{spin parameter} & a_* \\
& \text{deformation parameter} & \epsilon_3 \\ \hline
\text{observer} & \text{viewing angle} & i \\ \hline
\text{ion torus} & \text{dimensionless specific angular momentum} & \lambda \\
& \text{magnetic to total pressure ratio} & \beta \\
& \text{polytropic index} & n \\
& \text{energy density at the center} & \rho_c \\
& \text{electron temperature at the center} & T_c \\
\end{array} \nonumber
\ee
We note, however, that $\lambda$ depends also on the spacetime, because the numerical values of $l_{\rm max}$ and $l_{\rm min}$ depend on the background metric.

We consider a polytropic index $n = 3/2$, which is suitable for a non-relativistic gas with negligible radiation pressure. The pressure $P$ is the sum of the magnetic pressure $P_m$ and the gas pressure $P_{gas}$, which are supposed to have a constant ratio, so 
\be
P = P_m + P_{gas} \, , \quad P_m = \frac{B^2}{24 \pi} = \beta P \, .
\ee
$B$ is the intensity of the magnetic field. The gas is described by a 2-temperature plasma, so the gas pressure is the sum of the ion and electron contributions. The pressure $P$, the energy density $\rho$, and the electron temperature $T$ can be written as~\cite{straub,vincent} 
\be
P &=& K \rho^{5/3} \, , \\
\rho &=& \frac{1}{K^n} \Big[\left(K \rho_c^{1/n} + 1\right)^\omega - 1 \Big]^n \, , \\
T &=& T_c \left( \frac{\rho}{\rho_c} \right)^{1/n}  \, , 
\ee
where 
\be
K = \frac{1}{\rho_c^{1/n}} \exp \left(\frac{W_c - W_s}{n+1} - 1\right) \, .
\ee

With this set-up, we can compute the spectrum of an ion torus. We use the ray-tracing code described in Ref.~\cite{cfm2}. The observed specific flux is given by
\be
F_{\rm obs} (\nu_{\rm obs}) = \int I_{\rm obs} (\nu_{\rm obs}) d\Omega_{\rm obs} 
= \int g^3 I_{\rm e} (\nu_{\rm e}) d\Omega_{\rm obs} \, ,
\ee
where $I_{\rm obs}$ and $I_{\rm e}$ are, respectively, the specific intensity of the radiation as measured by the distant observer and in the rest frame of the accreting gas. $d\Omega_{\rm obs} = dX dY / d^2$ is the element of the solid angle subtended by the image of the disk on the observer's sky, $X$ and $Y$ are the Cartesian coordinates on the observer's sky, and $d$ is the distance of the BH from the observer. $I_{\rm obs} = g^3 I_{\rm e}$ follows from Liouville's theorem. $g$ is the redshift factor
\be
g = \frac{k_\mu u^\mu_{\rm obs}}{k_\nu u^\nu} =
\frac{1}{u^t}\left(1 + \Omega \frac{k_\phi}{k_t}\right)^{-1}
= \sqrt{- g_{tt} - 2 \Omega g_{t\phi} - \Omega^2 g_{\phi\phi}} 
\left(1 + \Omega \frac{k_\phi}{k_t}\right)^{-1} \, ,
\ee
where $k^\mu$ is the 4-momentum of the photon, $u^\mu_{\rm obs} = (1,0,0,0)$ is the 4-velocity of the distant observer, and $u^\mu$ is the 4-velocity of the fluid element. $k_\phi/k_t$ is a constant of motion along the photon path. The observer's sky is divided into a number of small elements and the ray-tracing procedure provides the observed flux density from each element; summing up all the elements, we get $F_{\rm obs} $.

Contrary to the case discussed in Ref.~\cite{cfm2}, here the disk is geometrically thick and optically thin. The specific intensity $I_{\rm e}$ is calculated by integrating along every photon path the emission and absorption contributions
\be
I_{\rm e} (\nu_{\rm e}) = 
\int \Big[j_{\rm e} (\nu_{\rm e}) - \alpha(\nu_{\rm e}) I_{\rm e} (\nu_{\rm e}) \Big] d\ell \, ,
\ee
where $j_{\rm e}$ and $\alpha$ are, respectively, the emission and absorption coefficient, $d\ell = u^\mu k_\mu d \tilde{\ell}$ is the infinitesimal proper length as measured in the rest frame of the emitter, and $\tilde{\ell}$ is the affine parameter of the photon trajectory. In what follows we assume $\alpha=0$. The electromagnetic spectrum of the ion torus is produced by bremsstrahlung, synchrotron processes, and inverse Compton scattering of both bremsstrahlung and synchrotron photons off free electrons in the medium. Creation and annihilation of electron-positron pairs are neglected. The details can be found in~\cite{straub,vincent} and references therein. The emission coefficient is thus the sum of four contributions
\be
j_{\rm e} (\nu_{\rm e}) = 
j_{\rm e}^{\rm brem} (\nu_{\rm e})
+ j_{\rm e}^{\rm sync} (\nu_{\rm e})
+ j_{\rm e}^{\rm Cbrem} (\nu_{\rm e})
+ j_{\rm e}^{\rm Csyn} (\nu_{\rm e}) \, .
\ee
The expression for these contributions are reported in~\cite{straub,vincent}.

\begin{figure}
\begin{center}
\includegraphics[type=pdf,ext=.pdf,read=.pdf,width=7.6cm]{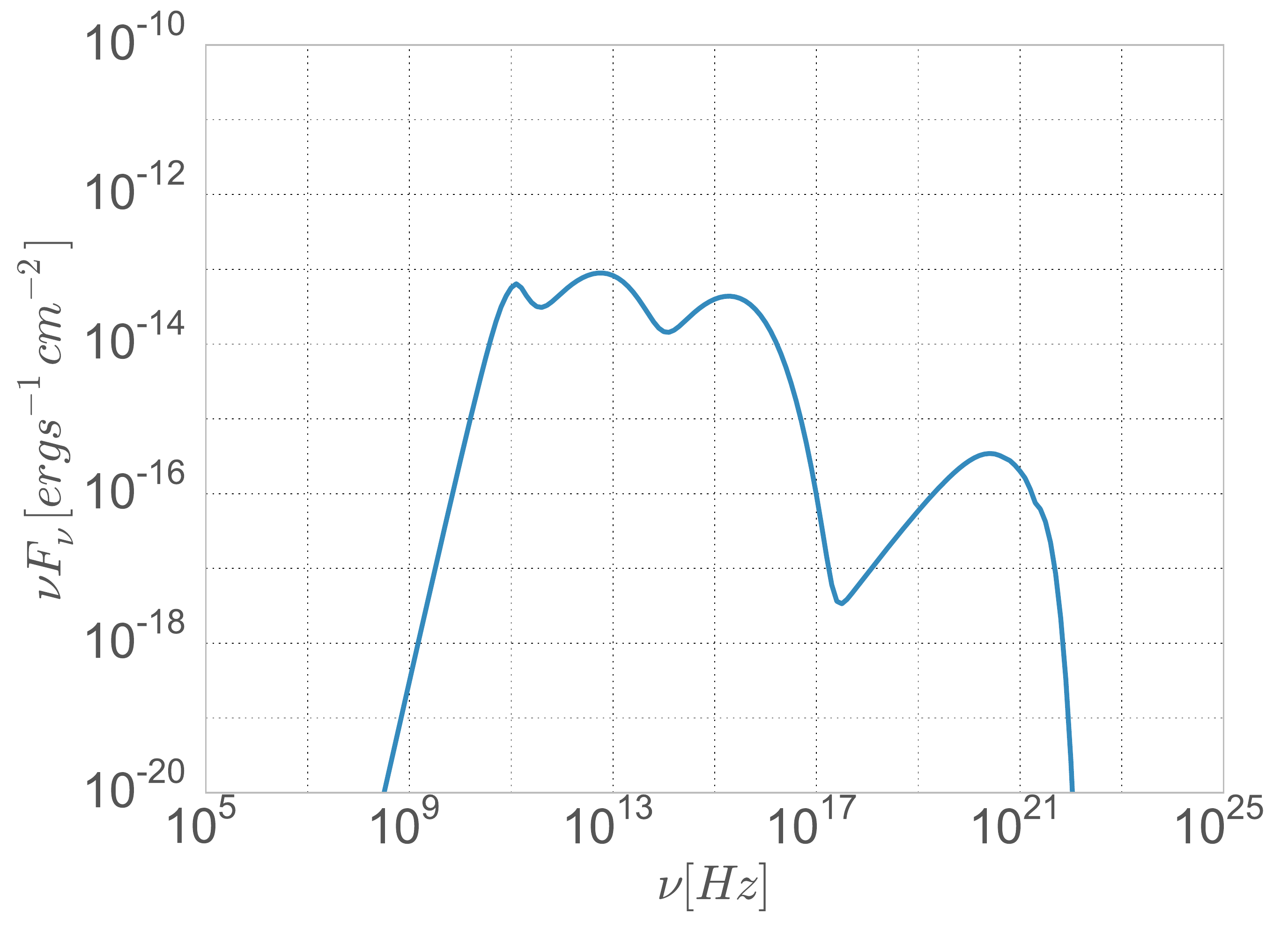}
\includegraphics[type=pdf,ext=.pdf,read=.pdf,width=7.6cm]{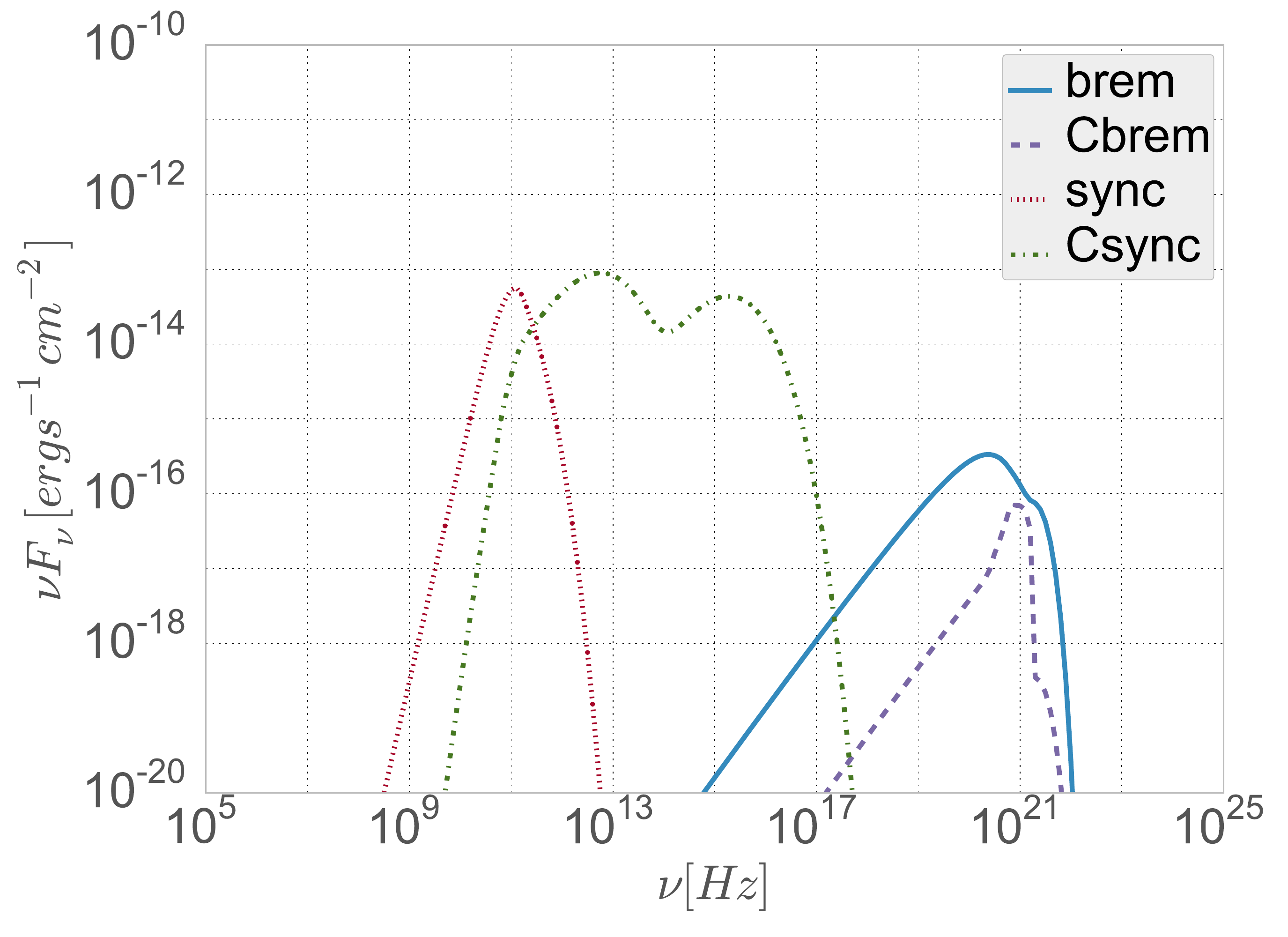}
\end{center}
\vspace{-0.5cm}
\caption{Left panel: spectrum of the ion torus of the default model ($a/M = 0.5$, $\epsilon_3 = 0$, $i = 60^\circ$, $\lambda=0.3$, $\beta=0.1$, $n=3/2$, $\rho_c = 10^{-17}$~g/cm$^3$, $T_c/T_v = 0.02$). Right panel: contribution from every radiative mechanism to the default model, namely bremsstrahlung emission (blue solid line), Comptonisation of bremsstrahlung emission (violet dashed line), synchrotron radiation (red dotted line), and Comptonisation of synchrotron radiation (green dashed-dotted line). See the text for more details.}
\label{fig1}
\end{figure}

\begin{figure}
\begin{center}
\includegraphics[type=pdf,ext=.pdf,read=.pdf,width=7.6cm]{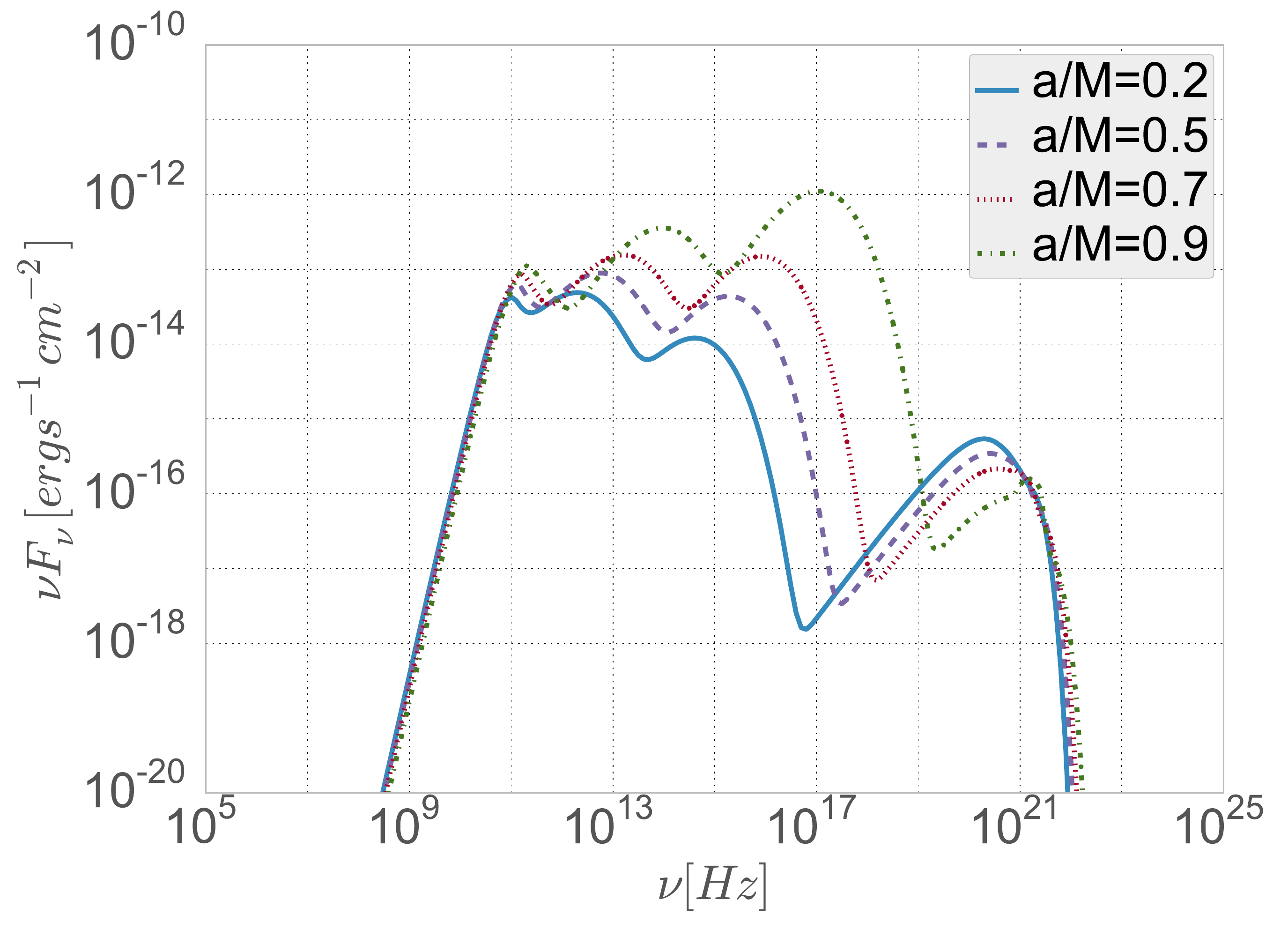}
\includegraphics[type=pdf,ext=.pdf,read=.pdf,width=7.6cm]{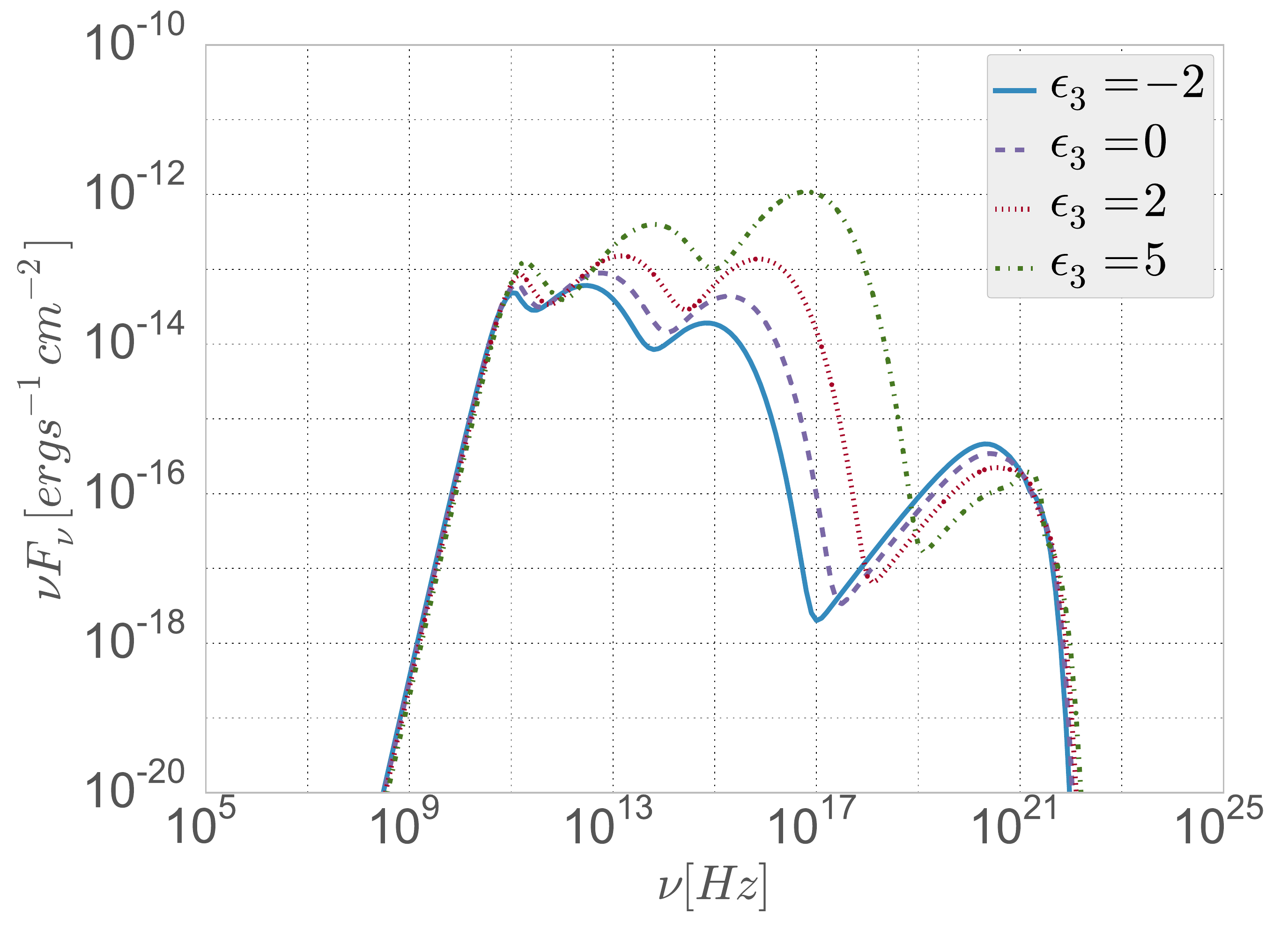}
\end{center}
\vspace{-0.5cm}
\caption{Effect of the spin parameter (left panel) and of the deformation parameter (right panel) on the spectrum of an ion torus. The other parameters are set to the value of the default model. See the text for more details.}
\label{fig2}
\vspace{0.8cm}
\begin{center}
\includegraphics[type=pdf,ext=.pdf,read=.pdf,width=7.6cm]{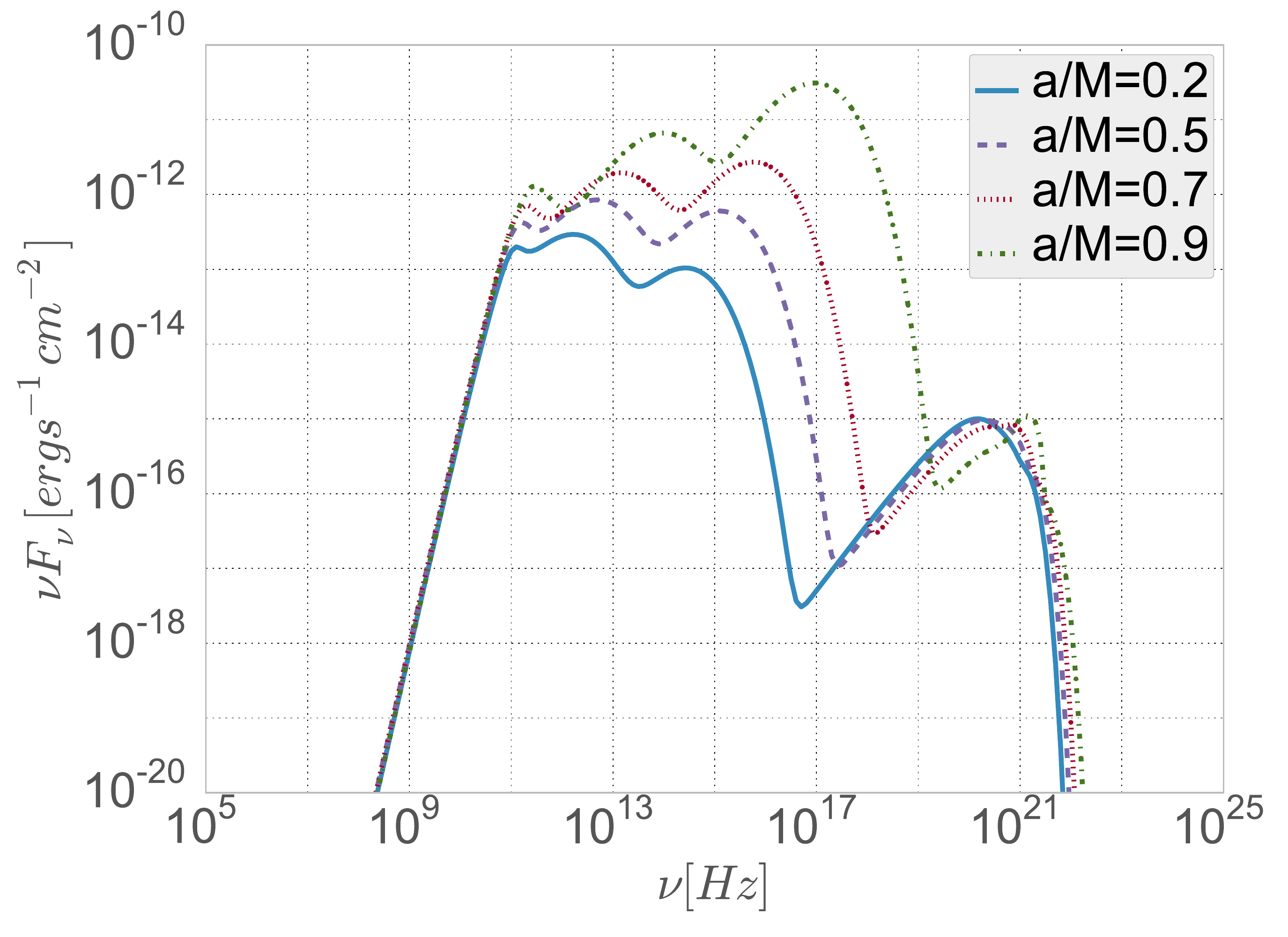}
\includegraphics[type=pdf,ext=.pdf,read=.pdf,width=7.6cm]{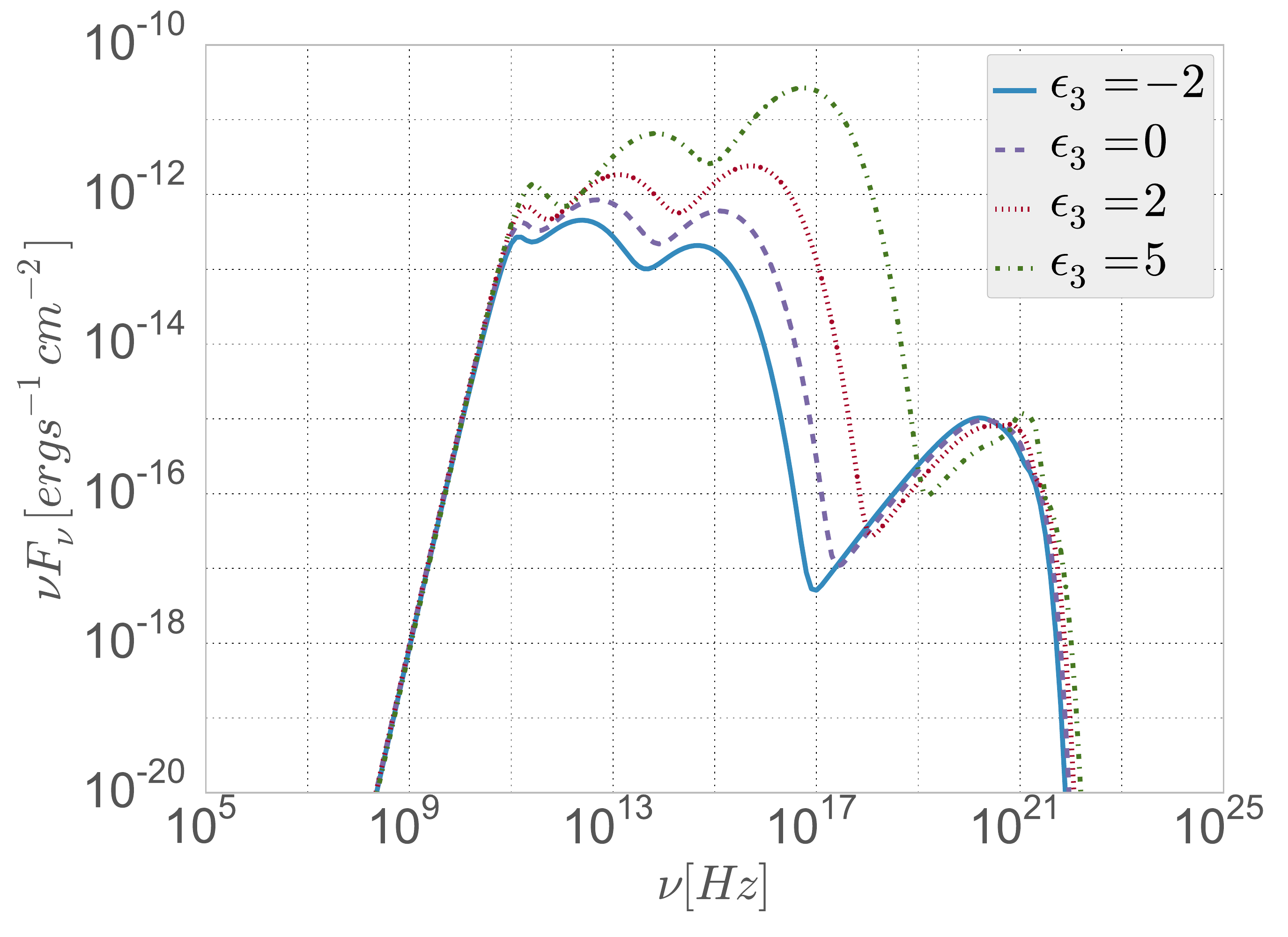}
\end{center}
\vspace{-0.5cm}
\caption{As in Fig.~\ref{fig2} for $\lambda=0.6$.}
\label{fig3}
\end{figure}

\begin{figure}
\begin{center}
\includegraphics[type=pdf,ext=.pdf,read=.pdf,width=7.6cm]{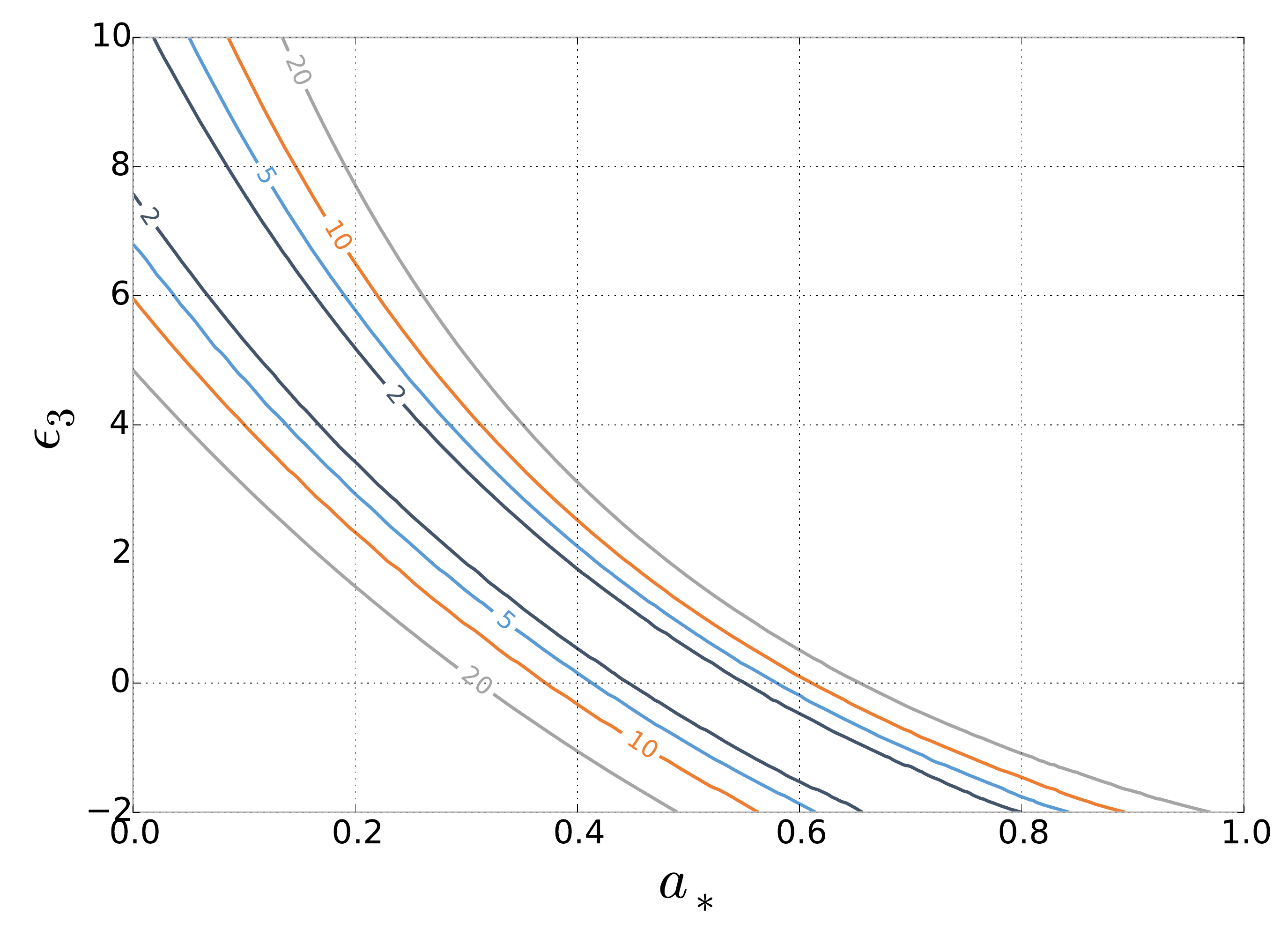}
\includegraphics[type=pdf,ext=.pdf,read=.pdf,width=7.6cm]{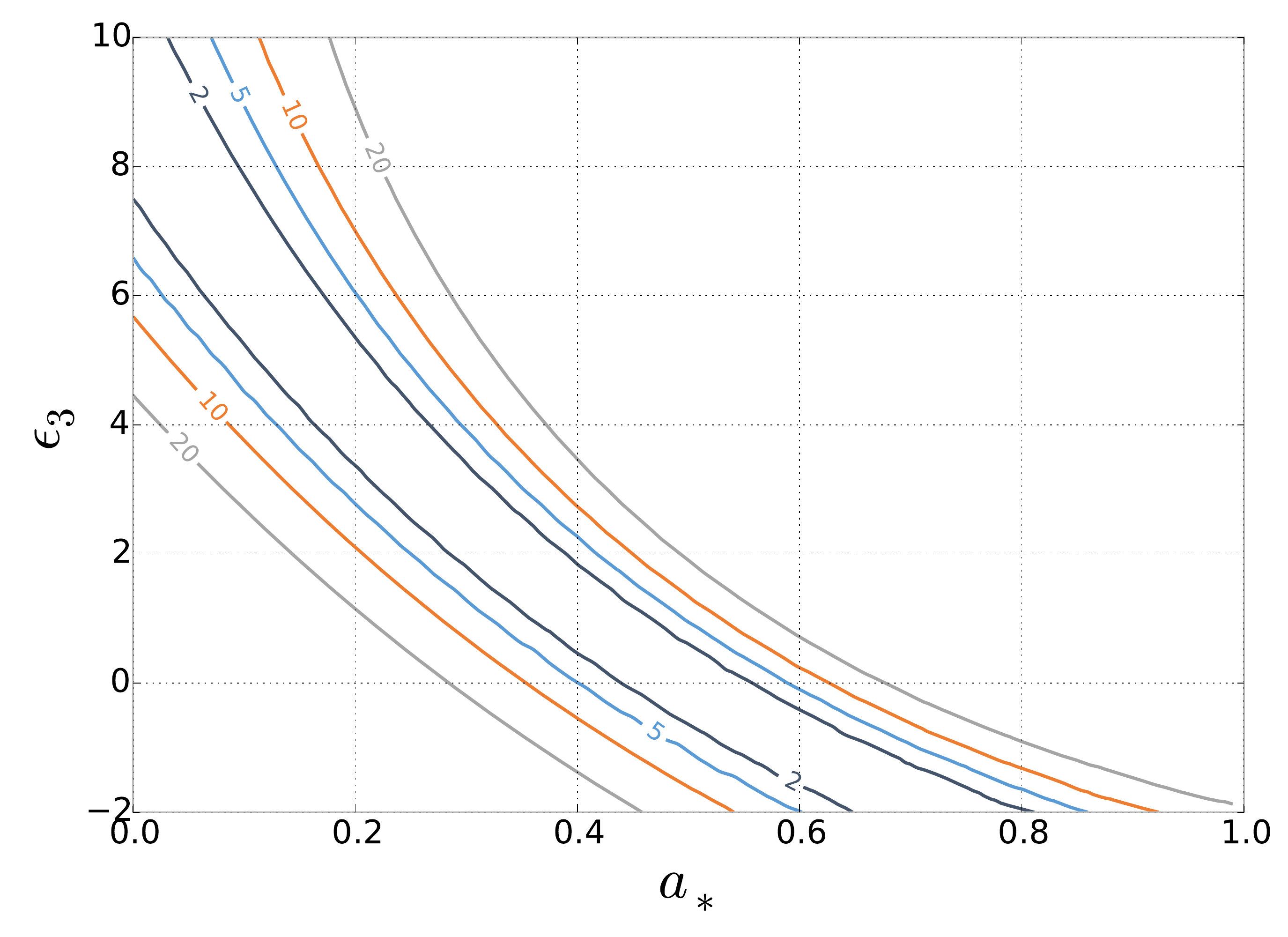}
\end{center}
\vspace{-0.5cm}
\caption{Contour map of the function $S$ in which the reference model has the parameters of the default model: $a_*^{\rm ref} = 0.5$, $\epsilon_3^{\rm ref} = 0$ (Kerr BH), $\lambda^{\rm ref}=0.3$, $\beta^{\rm ref}=0.1$, $n^{\rm ref}=3/2$, $\rho^{\rm ref}_c = 10^{-17}$~g/cm$^3$, and $T_c^{\rm ref}/T_v = 0.02$. In the left panel, $\lambda = 0.3$ is fixed, while in the right panel $\lambda$ is free and $S$ is minimized with respect to it. See the text for more details.}
\label{fig4}
\vspace{0.8cm}
\begin{center}
\includegraphics[type=pdf,ext=.pdf,read=.pdf,width=7.6cm]{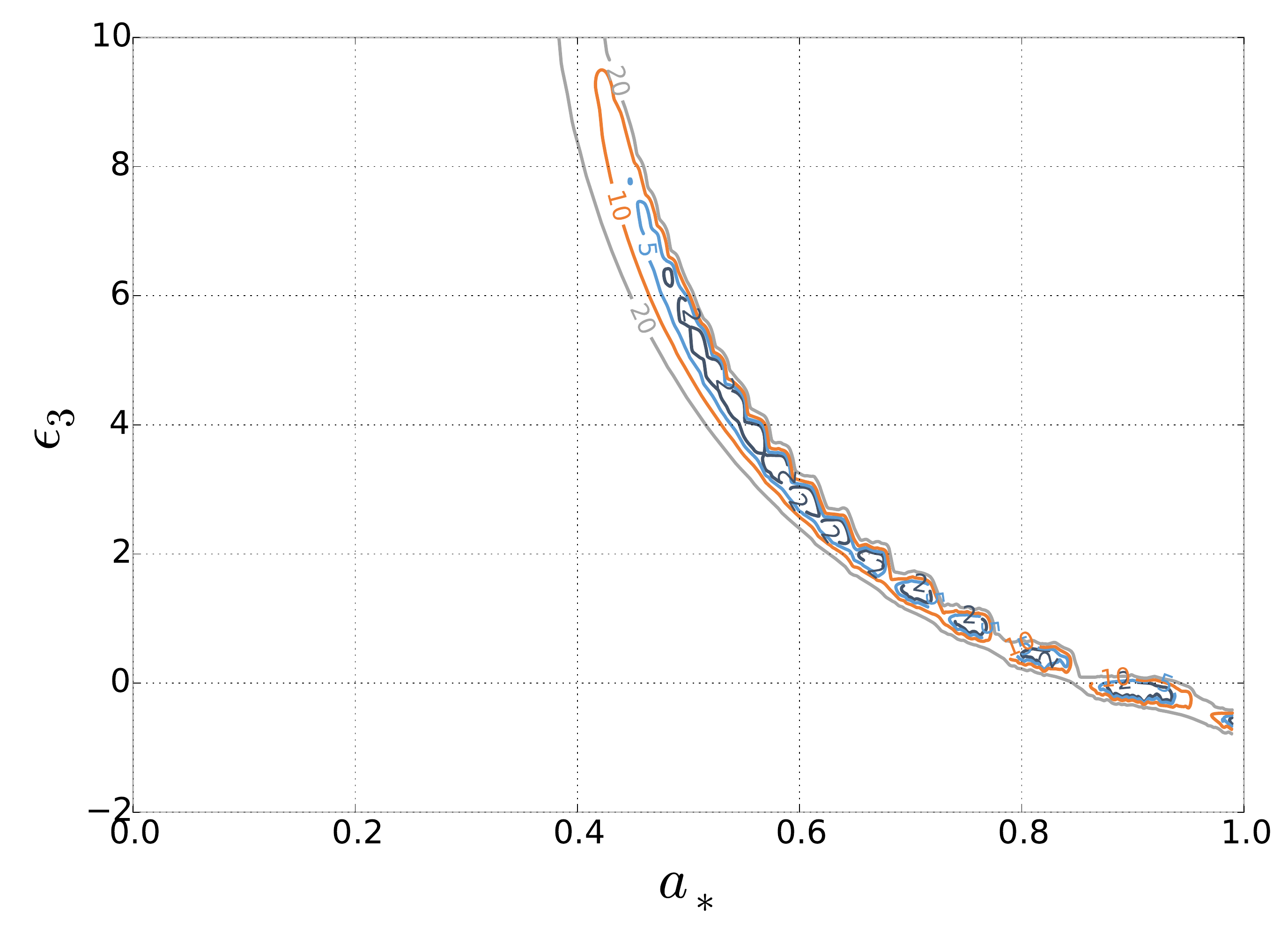}
\includegraphics[type=pdf,ext=.pdf,read=.pdf,width=7.6cm]{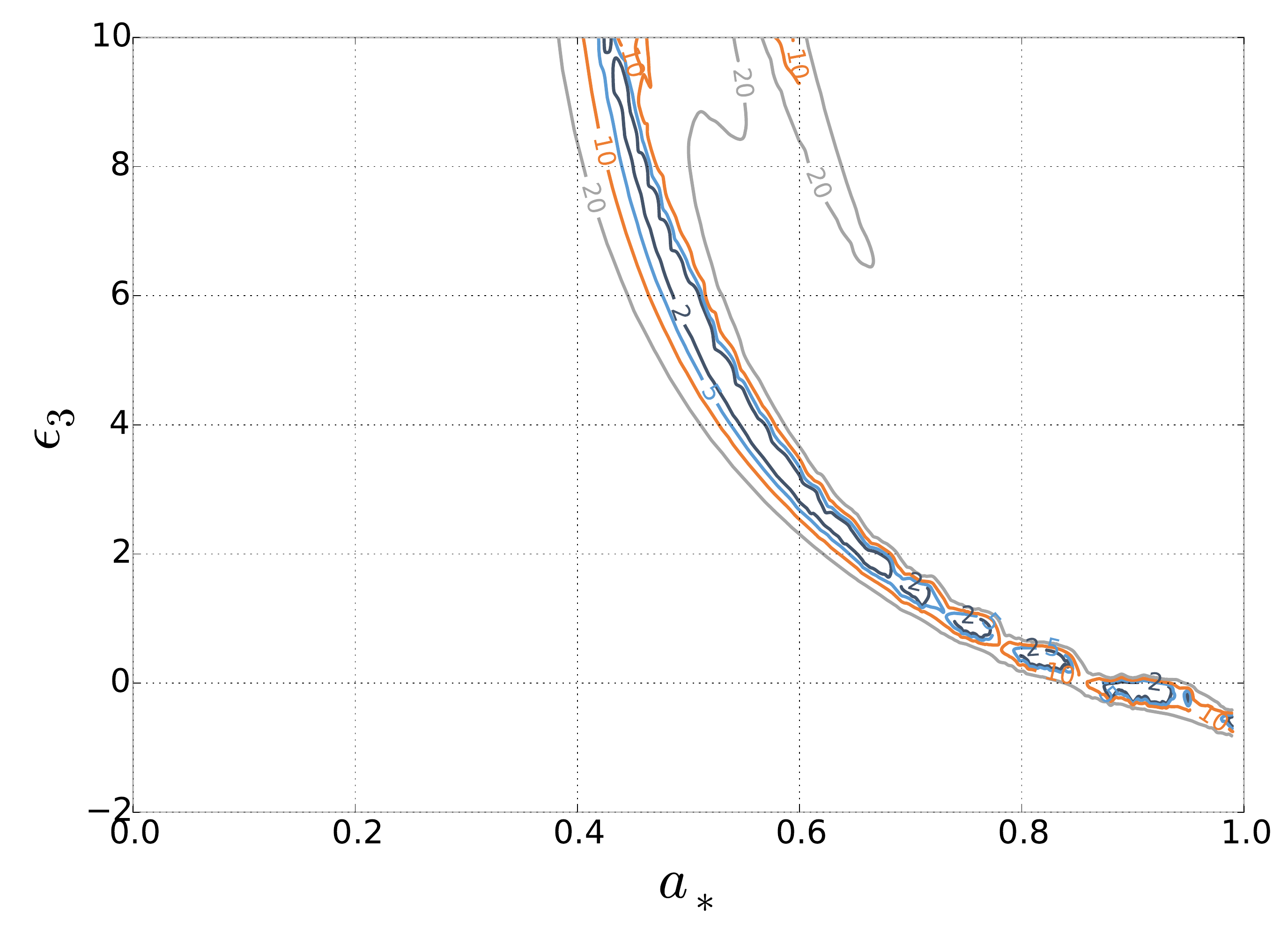}
\end{center}
\vspace{-0.5cm}
\caption{As in Fig.~\ref{fig4} with the reference model of $a_* = 0.9$ and $\epsilon_3 = 0$. In the left panel, $\lambda = 0.3$ is fixed, while in the right panel $\lambda$ is free and $S$ is minimized with respect to it. See the text for more details.}
\label{fig5}
\end{figure}

\section{Simulations \label{s-4}}

Fig.~\ref{fig1} show the spectrum of our default model. The background is described by the Kerr metric ($\epsilon_3 = 0$) with spin parameter $a_*=0.5$. The viewing angle is $i = 60^\circ$. Concerning the ion torus, we have assumed $\lambda=0.3$, $\beta=0.1$, $n=3/2$, $\rho_2 = 10^{-17}$~g/cm$^3$, and $T_c/T_v = 0.02$ where $T_v$ is the virial temperature. The left panel shows the total spectrum, while the right panel shows the contributions from the emission mechanisms under consideration (bremsstrahlung, synchrotron emission, and inverse Compton scattering of both bremsstrahlung and synchrotron photons).

The impact of the parameters of the model on the spectrum of the ion torus have been already discussed and illustrated in Ref.~\cite{straub}. Here we just show the effect of the spin parameter $a_*$ and of our new parameter $\epsilon_3$ describing possible deviations from the Kerr geometry. Figs.~\ref{fig2} and \ref{fig3} show some spectra for different values of $a_*$ and $\epsilon_3$, respectively in the case $\lambda=0.3$ and 0.6. As in the case of the thermal spectrum of thin disks (see e.g. Fig.~2 in Ref.~\cite{cb2}), the impact of $a_*$ and $\epsilon_3$ looks similar and the spectrum could be degenerate with respect to these two parameters. However, the spectrum of a ion torus is more complicated and the conclusion is not obvious.

To be a little bit more quantitative, we consider a reference model and we compare its spectrum with those produced in other models by varying the spin parameter, the deformation parameter, and possibly the specific fluid angular momentum $\lambda$. We use the following function as an estimator
\be
S(a_*, \epsilon_3, \lambda) = \sum_i \left[ \frac{\log F_i(a_*, \epsilon_3, \lambda) 
- \log F_i^{\rm ref}}{C \log F_i^{\rm ref}} \right]^2 \, .
\ee 
where $F_i^{\rm ref}$ is the observed intensity flux of the reference model at the frequency $i$, $F_i(a_*, \epsilon_3, \lambda)$ is the observed intensity flux of the model with $(a_*, \epsilon_3, \lambda)$, and $C$ is a constant that we have arbitrarily set to 0.1. All the other parameters are fixed to the value of the default model reported above.

In Fig.~\ref{fig4}, we show some contour levels of $S$ in the case the parameters of the reference model are set to the values of the default model, whose spectrum is shown in Fig.~\ref{fig1}. In the left panel, $\lambda = 0.3$ is fixed. In the right panel, $\lambda$ is a free parameter and $S$ is minimized with respect to it, as one should do if $\lambda$ is unknown and to be determined by the fit. These plots clearly show the strong degeneracy between $a_*$ and $\epsilon_3$: it is impossible to get an estimate of one of these parameters without knowing the other from an independent measurement. Even if under the assumption of the Kerr background we could obtain a good constraint on the spin like $0.45 < a_* < 0.55$ (the contour line $S = 2$), actually we cannot exclude large deviations from the Kerr background and a possibly very different value of the spin $a_*$. The impact of $\lambda$ is modest: when even this parameter is free, the constraints on $a_*$ and $\epsilon_3$ are roughly the same as in the case of a fixed $\lambda$. The degeneracy of the spectrum with respect to $a_*$ and $\epsilon_3$ is very similar to that found in the measurements of the thermal spectrum of thin disks, see e.g. Ref.~\cite{cfm3}.

Fig.~\ref{fig5} is like Fig.~\ref{fig4}, but now the reference BH has $a_* = 0.9$ (all the other parameters are unchanged). $\lambda$ is fixed in the left panel, while it is free in the right panel. As with other approaches, fast-rotating BHs are more suitable to test the Kerr metric and the constraints are stronger. However, the parameters $a_*$ and $\epsilon_3$ are still very correlated and it is impossible to rule out large deviations from the Kerr geometry.

\section{Concluding remarks \label{s-5}}

SgrA$^*$ is the supermassive BH candidate at the center of our Galaxy and a special source to test the Kerr metric. Unlike most of the other BH candidates, SgrA$^*$ may be soon studied with many different approaches and the combination of these measurements could provide an independent estimate of both the spin parameter and possible deviations from the Kerr solution. The general problem to test the Kerr metric is indeed that observational features of Kerr BHs may be reproduced by non-Kerr BHs with a different spin. If different measurements are sensitive to different relativistic effects, there is the possibility of breaking such a degeneracy and unambiguously testing the Kerr paradigm.

In this paper, we have explored the possibility of testing SgrA$^*$ with the spectrum of its accretion structure. Following Ref.~\cite{straub}, we have employed the Polish doughnut model to describe an optically thin ion torus radiating via bremsstrahlung, synchrotron processes, and inverse Compton scattering. The model has 8 free parameters: 2 parameters related to the spacetime metric (spin parameter $a_*$, deformation parameter $\epsilon_3$), 1 parameter associated to the observer (viewing angle $i$), and 5 parameters to describe the ion torus (dimensionless specific angular momentum $\lambda$, magnetic to total pressure ratio $\beta$, polytropic index $n$, energy density at the center $\rho_c$, and electron temperature at the center $T_c$).

With the use of the ray-tracing code of Ref.~\cite{cfm2}, we have computed the spectrum of an ion torus with specific values of these 8 parameters. In the presence of good data of the spectrum of the accretion structure of SgrA$^*$, one could potentially fit the observations and constrain the parameters of the model. In this explorative work, we have limited our attention to the spin, the deformation parameter, and to the fluid angular momentum, but a more accurate study should also check the correlation with the other parameters of the model. As in the case of the thermal spectrum of thin accretion disks, the spectrum of an ion torus is degenerate with respect to $a_*$ and $\epsilon_3$ and it is impossible to determine one of the two parameters without an independent measurement of the other. In this sense, the sole observation of the spectrum of an ion torus cannot test the Kerr metric, even in the presence of high quality data and with the systematics under control. However, in the case of SgrA$^*$ it will be hopefully possible to combine several measurements. The approach studied in this paper may still provide interesting results if its constraints could be combined with those from other observations, and in particular from the observation of the BH shadow and from the detection of a pulsar orbiting with a period of a few months.


\begin{acknowledgments}
This work was supported by the NSFC grant No.~11305038, the Shanghai Municipal Education Commission grant for Innovative Programs No.~14ZZ001, the Thousand Young Talents Program, and Fudan University.
\end{acknowledgments}


\end{document}